# Atomic-Number (*Z*)-Correlated Atomic Sizes for Deciphering Electron Microscopic Molecular Images

Junfei Xing, Keishi Takeuchi, Ko Kamei, Takayuki Nakamuro, Koji Harano,* Eiichi Nakamura*

Department of Chemistry, The University of Tokyo, 7-3-1 Hongo, Bunkyo-ku, Tokyo 113-0033, Japan

*Correspondence to: harano@chem.s.u-tokyo.ac.jp, nakamura@chem.s.u-tokyo.ac.jp

**Abstract**

With the advent of atomic-resolution transmission electron microscopy (AR-TEM) achieving sub-Ångstrom image resolution and submillisecond time resolution, an era of visual molecular science where chemists can visually study the time evolution of molecular motions and reactions at atomistic precision has arrived. However, the appearance of experimental TEM images often differs greatly from that of conventional molecular models, and the images are difficult to decipher unless we know in advance the structure of the specimen molecules. The difference arises from the fundamental design of the molecular models that represent atomic connectivity and/or the electronic properties of molecules rather than the nuclear charge of atoms and electrostatic potentials that are felt by the e-beam in TEM imaging. We found a good correlation between the atomic number (*Z*) and the atomic size seen in TEM images when we consider shot noise in digital images. We propose here *Z*-correlated (ZC) atomic radii for modeling AR-TEM images of single molecules and ultrathin crystals, with which we can develop a good estimate of the molecular structure from the TEM image much more easily than with conventional molecular models. Two parameter sets were developed for TEM images recorded under high-noise ($ZC_{HN}$) and low-noise ($ZC_{LN}$) conditions. The new molecular models will stimulate the imaginations of chemists planning to use AR-TEM for their research.

**Main Text**

The world of atoms and molecules invisible to human eyes has been illuminated by the development of three-dimensional models, in which atoms represented by spheres are connected to each other with appropriate spatial disposition. Here, the choice of the radii of the spheres, in addition to coloring, is the most crucial visual identifier of atoms and atomic ions, and a variety of useful constructions of radii, either empirical or quantum mechanical,[1,2] have been proposed in accord with chemical intuition. Whereas wire and ball-and-stick models are primitive, showing only bond lengths, bond angles, and torsional angles, chemists have implemented further information into molecular models. Representing the van der Waals isosurface, a Corey–Pauling–Koltun (CPK) model was developed to illustrate the molecular surface for analysis of intermolecular interactions.[3] Shannon's ionic model[4] is useful for inorganic chemistry. The atomic radii chosen in these models reflect the electronic properties of atoms and molecules.

With the advent of AR-TEM achieving sub-Ångstrom image resolution and submillisecond time resolution,[5,6,7,8] an era of "visual molecular science" has arrived, where chemists can visually study the time evolution of molecular motions and reactions at atomistic precision.[9,10,11] However, experimental TEM images often differ greatly from what chemists expect using their favorite molecular models (cf Figure 1), and this discrepancy contributed to reducing the popularity of AR-TEM in molecular science. The difficulty is particularly important when we deal with unknown structures that are mobile and lack structural periodicity. For example, structure assignment of the AR-TEM images of a rotating zinc cluster (Figure 1a) that forms in the synthesis of a crystal of a metal–organic framework (MOF; MOF-5)[12] was impossible without reducing the number of initial guess structures down to several hundred from a total of 4096 isomers and their rotamers as to the 2-iodo-1,4-phenylene bridges (Figure 1d, e). Here, we found the CPK atomic radii (Figure 1b) more confusing than revealing and became convinced of a need for new atomic radii that better represent the experimental images (compare Figure 1b and 1c). In doing so, we found that the atom size seen in TEM images shows a strong correlation to the atomic number ($Z$), particularly if we consider the shot noise in digital images. In a noise-free image, we find little correlation, as theory predicts.[13] Our idea may be symbolically described as comparing the sizes of mountains by measuring the width of the lowest portion of each mountain that is visible above a sea of clouds. We report here a set of $Z$-correlated (ZC) atomic radii for deciphering AR-TEM images, with which we can develop a good estimate of the molecular structure from the TEM image much more easily than with conventional molecular models. Two parameter sets are described: one for TEM images under high-noise conditions ($ZC_{HN}$, Figure 2c), and another under low-noise conditions ($ZC_{LN}$, Figure 2d). The former is useful for single-molecule imaging and the latter for an ultrathin film such as UiO-66 MOF (Figure 1f).[14,15] The ZC radii reflect the relative sizes of various elements and help chemists develop an initial estimate of the molecular structures. Note, however, that we do not intend the ZC model to reproduce the TEM images exactly. No chemists expect the CPK model to represent the van der Waals properties of the molecule exactly, and molecular models are there to stimulate chemists' imaginations.

In Figure 2, we compare the atomic radii used in the CPK model, Shannon's ionic model, and the $ZC_{LN}$ and $ZC_{HN}$ models. CPK radii are smaller for transition metals and larger for main group elements, and Shannon's radii decrease with the increasing atomic number within the same period. These atomic radii are unsuitable for deciphering AR-TEM images, where the behavior of the e-beam in the electrostatic potential created by atomic nuclei plays a decisive role.[16,17] In the ZC model proposed here, the atomic radius shows element dependence, changing systematically as $Z$ increases. The $ZC_{HN}$ atomic radii (Figure 2c) are smaller in signal size than the $ZC_{LN}$ radii (Figure 2d), show large $Z$-dependence, and are suitable for fast molecular imaging where the level of shot noise is large (e.g., Figure 1a–c).[18] The $ZC_{LN}$ radii are suitable for images of ultrathin crystalline films where the level of shot noise is comparatively low. The atomic sizes for the graphical construction of the ZC models are listed in Figure 2e. We anticipate that the new molecular model will stimulate the imaginations of chemists planning to use AR-TEM for their research.

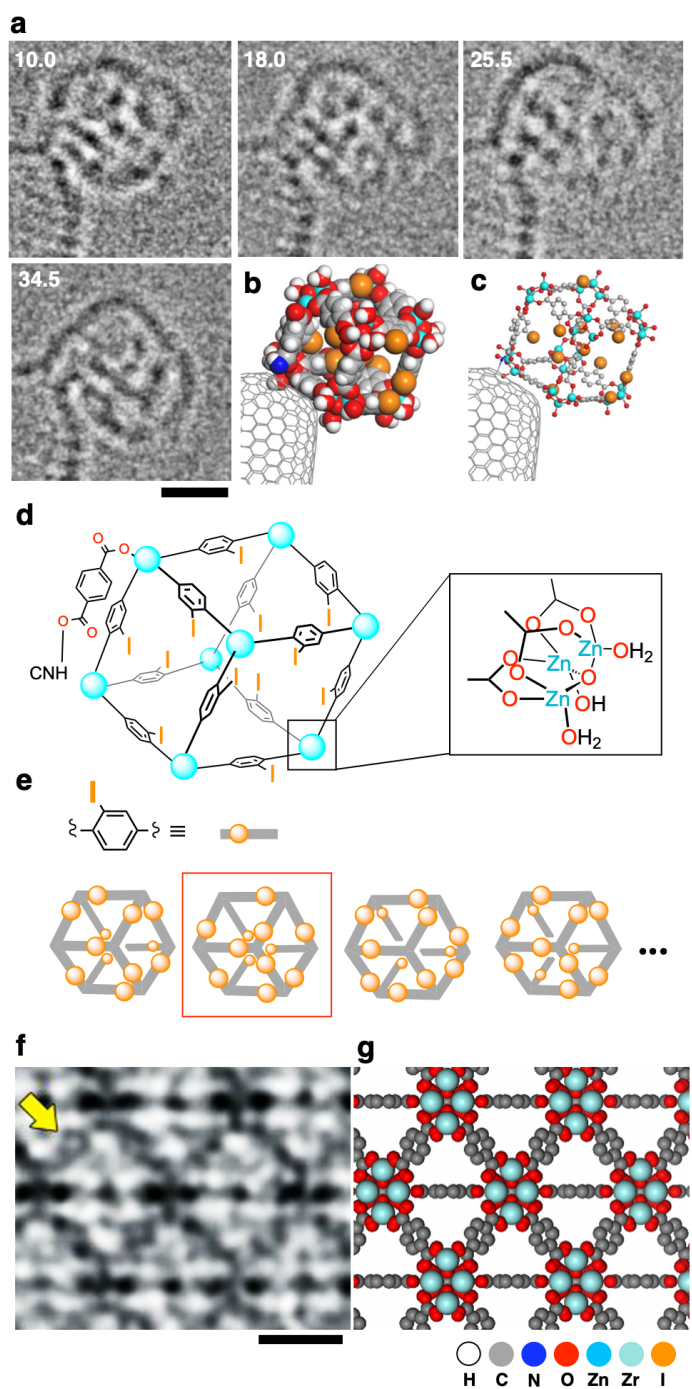

**Figure 1.** Experimental AR-TEM image and atomic-number-correlated molecular model (ZC model). (a) Time evolution of TEM image of a MOF-5 precursor, a cubic cluster made of $Zn^{2+}$ and 2-iodoterephthalic acid captured on a carbon nanotube. The molecular images are blurred by rotation and vibration. Time is shown in seconds. Taken from ref. 12. Copyright 2019 Nature Publishing Group. (b) A CPK model

corresponding to the image at 10.0 s. (c) A ZC$_{HN}$ model also for the image at 10.0 s. (d) Chemical structure of the cubic cluster. (e) Schematic illustration of isomers of the cubic MOF-5 precursor. Four structures out of 4096 possible isomers are shown. The structure seen in Figure 1a is highlighted by a red square. (f) TEM image of an ultrathin crystal of UiO-66 (taken from ref. 15). The benzene ring in a benzene dicarboxylate linker is indicated by an arrow. Copyright 2018 The American Association for the Advancement of Science. (g) A ZC$_{LN}$ model corresponding to f. Scale bars: 1 nm.

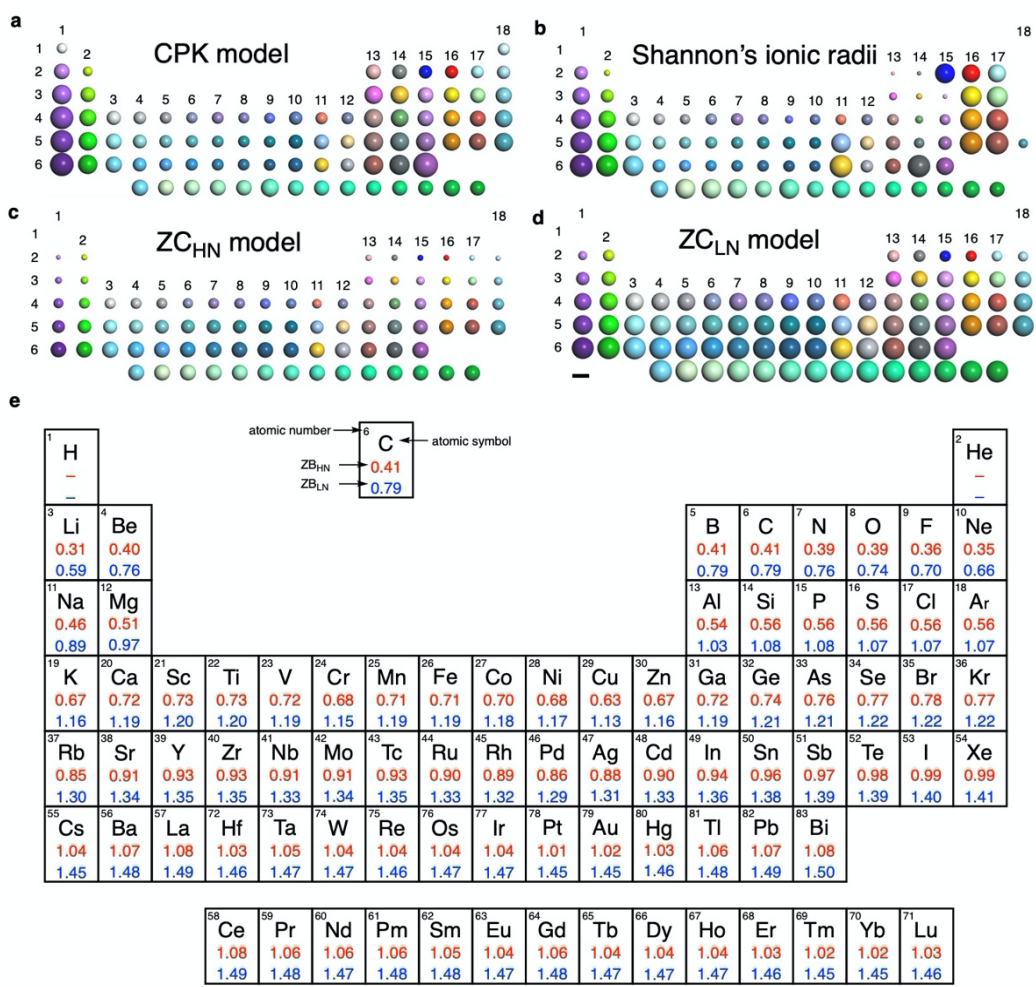

**Figure 2.** Periodic table of atomic sizes used in molecular models. (a) CPK model. (b) Shannon's ionic parameters. (c) ZC$_{HN}$ model. (d) ZC$_{LN}$ model. Scale bar: 2 Å. (e) List of ZC$_{HN}$ radii for single-molecule imaging and ZC$_{LN}$ radii for ultrathin crystals (in Å).

**Results and Discussion**

Experimental study of individual molecules at an atomistic level has been a challenge for chemists. Scanning probe microscopy is largely limited to imaging static specimen molecules on a flat substrate.[19,20] Utilizing specimens attached to a substrate or those comprising a part of a large crystal, the conventional TEM technique has suffered from similar limitations. AR-TEM imaging of single molecules in or on a carbon nanotube (CNT)[10] or on graphene[21,22] has realized dynamic molecular imaging against a vacuum background, hence opening an avenue to single-molecule atomic-resolution time-resolved electron microscopic (SMART-EM) imaging.[23] Here, we can study the time evolution of structural changes and reactions with up to millisecond speed and sub-Å spatial resolution.[18] The method produces a series of two-dimensional projection images where information on three-dimensional locations and properties of atomic nuclei is hidden (Figure 1a). Here, we see several dense and large signals, and less dense and smaller signals, ascribable to iodine and light atoms, respectively. As applied to a MOF named UiO-66, TEM provided an image depicting the zirconium atom as both dense and very large spots, and benzene rings as faint and small signals (yellow arrow, Figure 1f). We felt that this density/size correlation found in these TEM images does not match the theory because the signal size should be independent of $Z$,[13] while the signal intensity of individual atoms should depend on $Z$,[16,24] empirically, $Z^{2/3}$.[25,26] In the present study, we found that the signal size does show significant $Z$-dependence if we take into account the shot noise present in every TEM image.[27] We exploited this dependence in defining the atom radii for the ZC model.

For molecules and molecularly thin objects, TEM images largely comprise the phase contrast created by interference between the transmitted wave and the wave scattered by the atoms.[16] We generated theoretically simulated TEM images using the multislice simulation software elbis incorporating a shot noise calculation algorithm developed by Hosokawa.[28] We studied commonly encountered elements between $Z = 1$ and 83 with defocus value (df) = –10 nm and acceleration voltage of 80 kV, which are the conditions typical for imaging molecules and their assemblies.[29,30,31,32,33] The atomic radii in the $ZC_{LN}$ and $ZC_{HN}$ models are defined as described in the following paragraphs.

In Figure 3a, we show the phase contrast without shot noise for group 14 elements placed 20 Å away from each other. The intensity profile consists of a triangular peak flanked by a series of fringe signals and the height of the triangle increases as $Z$ increases, roughly proportional to $Z^{2/3}$ (Figure 3b, see Figure S1 for simulation images of all the elements).[25,26] The small undulation of the signal height for each period reflects the $Z$-dependence of nuclear charge.[13]

The base of the isosceles triangle, on the other hand, shows little $Z$-dependence (purple arrows, Figure 3a) as reported previously,[13] which led us to contemplate the origin of the experimentally observed $Z$-dependence of the image sizes. Some trial-and-error studies revealed that the length of the bottom of the triangle shows $Z$-dependence when the triangle is truncated by background noise. An intensity profile of the single-atom images overlaid with a background noise corresponding to an electron dose (ED) of 4.5 × $10^6$ e$^-$ nm$^{-2}$ frame$^{-1}$ is shown in Figure 3c, where we see an obvious effect of noise on the apparent signal height and also on the apparent radius of the signal, the noise acting like the sea of clouds that truncates the view of mountains. Thus, we set the standard deviation of the background noise as the threshold for visual recognition.[34] Cutting the triangle of the noise-free atom signals at low and high noise levels (LN and HN, blue and red lines in Figure 3a) produces smaller isosceles triangles, and we use the length of the bottom of these triangles (image radius in Figure 3e) as the atomic radii of the $ZC_{LN}$ and $ZC_{HN}$ models shown in Figure 2e. The blue line represents the LN conditions of an ED of $4.5 \times 10^6$ $e^-$ $nm^{-2}$ $frame^{-1}$ and the red line the HN conditions of an ED of $5.0 \times 10^5$ $e^-$ $nm^{-2}$ $frame^{-1}$—a typical dose used for fast molecular imaging using a CMOS detector.[12] In Figure 3d, we illustrate simulations of atomic images without shot noise and with low and high noise, to which the $ZC_{HN}$ and $ZC_{LN}$ models compare favorably (see Figures S2 and S3 for simulated images under LN and HN conditions for all the elements).

In Figure 3e, we summarize graphically the image radii calculated by truncating the bottom of the atomic signal with three noise threshold values (no, low, and high noise level) for all elements for $Z = 1$ to 83. A small undulation due to variation of effective nuclear charge is also seen. Numerical data on the ZC atomic radii are summarized in Figure 2e and the atomic images in Figure 2c and 2d. In our previous report,[12] we empirically found a correlation between $Z$ and apparent atomic radii, and here we propose the ZC model on a solid and broad basis. Although the variation of df (–4 to –15 nm) systematically changes the image radii by less than 20%, the trend of element dependence of the radius still holds (Figure 3f).[35] Though TEM images may be blurred for various reasons such as specimen motions, energy spread of the e-beam, defocus, and pixel, they contribute equally to all elements and hence does not much affect the *Z*-dependence of the image radii. Technical improvement will reduce the magnitude of blurring in the near future.

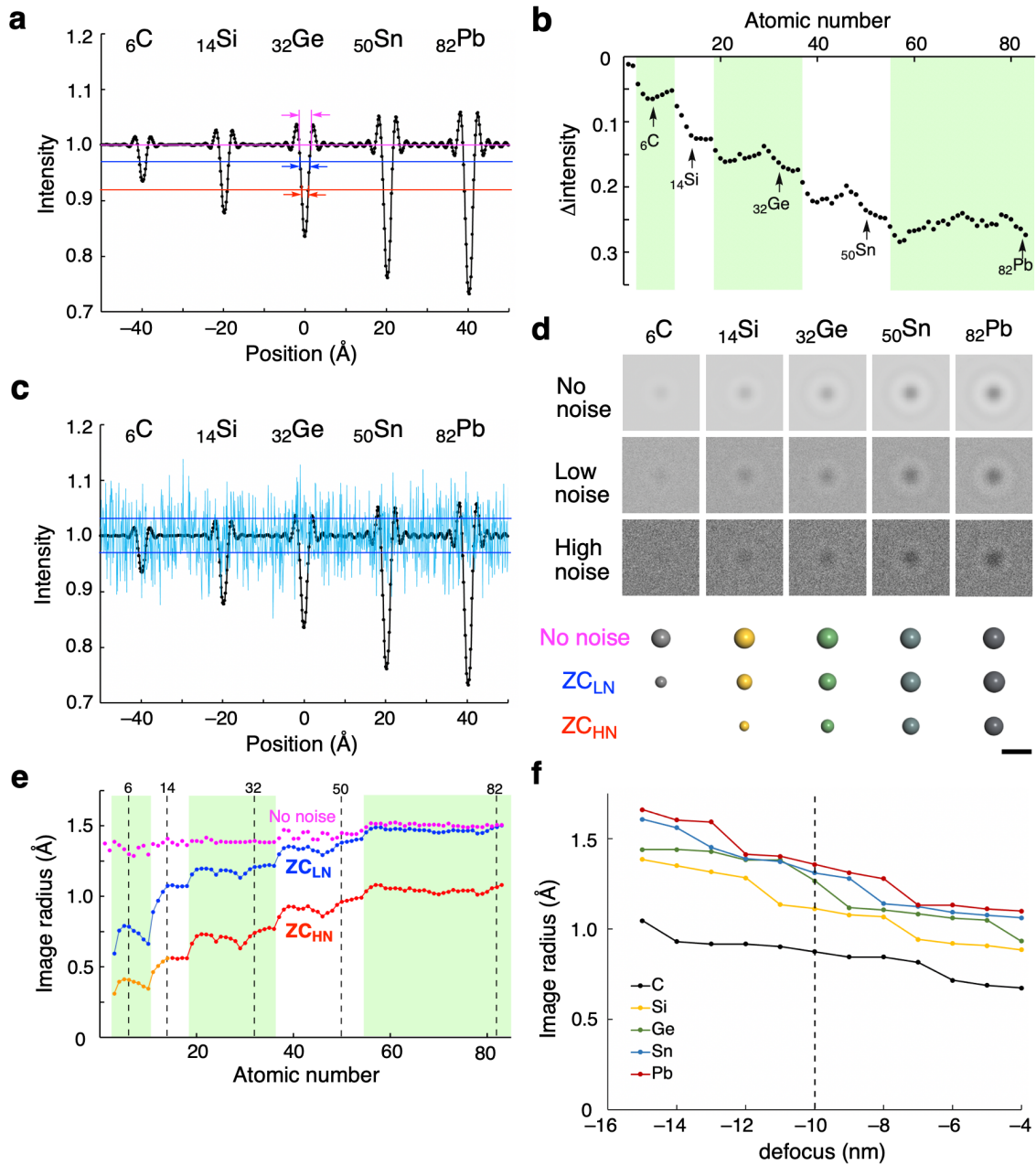

**Figure 3.** Image size of a single atom in simulated TEM images (80 kV, defocus = –10 nm). (a) Intensity profile of simulated images without noise of carbon, silicon,

germanium, tin, and lead atoms (i.e., images taken at an infinitely large ED). The intensity is normalized to the background intensity as 1 (purple line). Blue and red lines show a noise level corresponding to $4.5 \times 10^6$ e$^-$ nm$^{-2}$ frame$^{-1}$ (LN condition, maximum ED available in the software), and $5.0 \times 10^5$ e$^-$ nm$^{-2}$ frame$^{-1}$ (HN condition), respectively. (b) Intensity difference between a peak top and the background for a single atom. Periods 2, 4, and 6 are highlighted by light green. (c) Intensity profile of the single-atom images shown together with low-level shot noise (shown as light blue). The noise level is shown as standard deviation (blue line as in a). (d) Simulated images under no-noise, LN, and HN conditions. Corresponding atom models. Scale bar: 5 Å. (e) $Z$-dependence of the no-noise (purple), ZC$_{HN}$ (blue), and ZC$_{LN}$ (red) atomic radii. For atoms smaller than $Z = 14$ (orange line), we scale down the ZC$_{LN}$ sizes because the images are so faint that we could not determine the signal size. The TEM simulation software elbis outputs the data as pixel images, and hence the data are noncontinuous. (f) Defocus-dependence of ZC$_{LN}$ image radii shown for group 14 elements.

Signal height and the signal radius truncated by shot noise level increase as several atoms are overlapped along the axis of an e-beam.[36] Figure 4a shows the intensity profile of simulated images of five hypothetical cumulene molecules placed along the e-beam, whereas Figure 4b shows the simulated images. As we impose a shot noise of $5.0 \times 10^5$ e$^-$ nm$^{-2}$ frame$^{-1}$ (red line), single C and C–C placed perpendicular to the e-beam become hidden behind the noise, while larger molecules remain to be seen above the background noise (Figure 4c).[37] A similar effect of atom overlap is observed also for nitrogen and oxygen. In Figure 1a, we see the effect of overlapping carbon and oxygen atoms in the outline of the zinc cluster molecule.

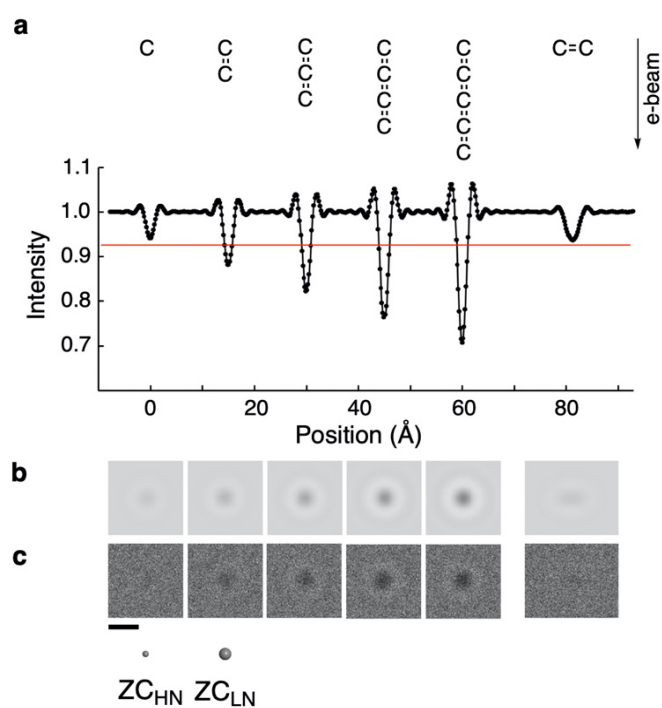

**Figure 4.** Simulated TEM images of overlapping carbon atoms. (a) Intensity profile of hypothetical cumulene molecules. The red line shows a noise level corresponding to an ED of $5.0 \times 10^5$ e$^-$ nm$^{-2}$ frame$^{-1}$ (high noise threshold in Figure 3). (b) Simulated TEM images under the no-noise condition. (c) Simulated images under the HN condition. Atom sizes of carbon for $ZC_{HN}$ and $ZC_{LN}$ models are also shown. Scale bar: 5 Å.

In Figure 5, we show several applications of the ZC model to SMART-EM imaging of organic and metallic molecular entities, where we use the $ZC_{HN}$ parameters and CPK model as a reference. Images in Figure 5a and c were taken on an instrument without aberration correction with a CCD camera at 120 kV, the one in Figure 5e on an aberration-corrected TEM with a CCD camera at 120 kV, and those in Figure 5b, d, f, and g on an aberration-corrected TEM equipped with a CMOS camera at 80 kV. In all cases, the $ZC_{HN}$ model that has been designed to represent atomic nuclei looks much closer to the TEM image than the space-filling model that represents the molecular surface.

Figure 5a illustrates a fullerene molecule bearing a perfluoroalkyl chain.[38] The fullerene moiety is elliptical because of vibration of the molecule along the CNT axis during imaging (0.5 s frame$^{-1}$). Figure 5b shows α-cyclodextrin (α-CD), a cyclic hexamer of glucose, sliding near the apex of a carbon nanohorn (CNH).[7] The ZC model matches the TEM image better than the CPK model. The larger size of individual dark spots in the TEM image than the ZC atoms is due to blurring caused by motions of the α-CD molecule during the 0.5 s exposure time.[7] Figure 5c shows a TEM image (0.5 s frame$^{-1}$) of a structurally mobile van der Waals complex involving 1,3,5-tris(4-bromophenyl)benzene,[39] and the mobility is supported by the broad TEM image (blurring) as compared with the more compact image of the ZC model. The location of eight bromine atoms can be more easily identified by the ZC model than by the CPK model. Figure 5d illustrates a trinuclear molybdenum complex attached to a hole on the graphitic surface of a CNH.[40] The crucial three molybdenum atoms are visible in the ZC model, while they are hidden behind oxygen atoms in the CPK model. Figure 5e shows an example of a TEM image of thyroxine, an amino acid containing four iodine atoms. The molecule interacts with the ammonium ion moiety on the tip of a CNT at 6.0 s, and slides gradually downward during 54.0 s, accompanying conformational change around the flexible ether linkage.

The advantage of the ZC model over conventional models such as wire, ball-and-stick, and CPK models is best illustrated with transition metal clusters. Figure 5f shows a TEM image of a $Au_{24}$ cluster on amorphous carbon, where the locations of the metal atoms are clearly discernable.[41] In an image of a $[CoSiW_{11}O_{39}]^{6-}$ polyoxometalate cluster in Figure 5g, we find only Co and W metal atoms. Accordingly, the $ZC_{HN}$ model shows only the heavy metals, while the CPK model shows only the oxygen atoms on the molecular surface. Note that we used the $ZC_{HN}$ radii because the background of the carbonaceous substrate acts as background noise in the TEM image.

In summary, the $ZC_{HN}$ atomic parameters reflecting nuclear charges illustrate the features of the TEM images better than the space-filling model reflecting van der Waals surfaces. Wire and ball-and-stick molecular models are not better than the space-filling model. Outlines of the CNTs appear thick in the TEM images due to the overlapping of carbon atoms as shown in Figure 4 using cumulene as an example, and also because the CNTs are vibrating faster than the TEM imaging rate.[18,42] Note that blur is caused by various factors in single-molecule TEM images. Although some factors such as defocus blur and energy spread of an e-beam are essentially unavoidable, a pixel blur caused by camera scintillation and a motion blur originating from a slow shutter speed will be minimized by improvement of the electron detector.

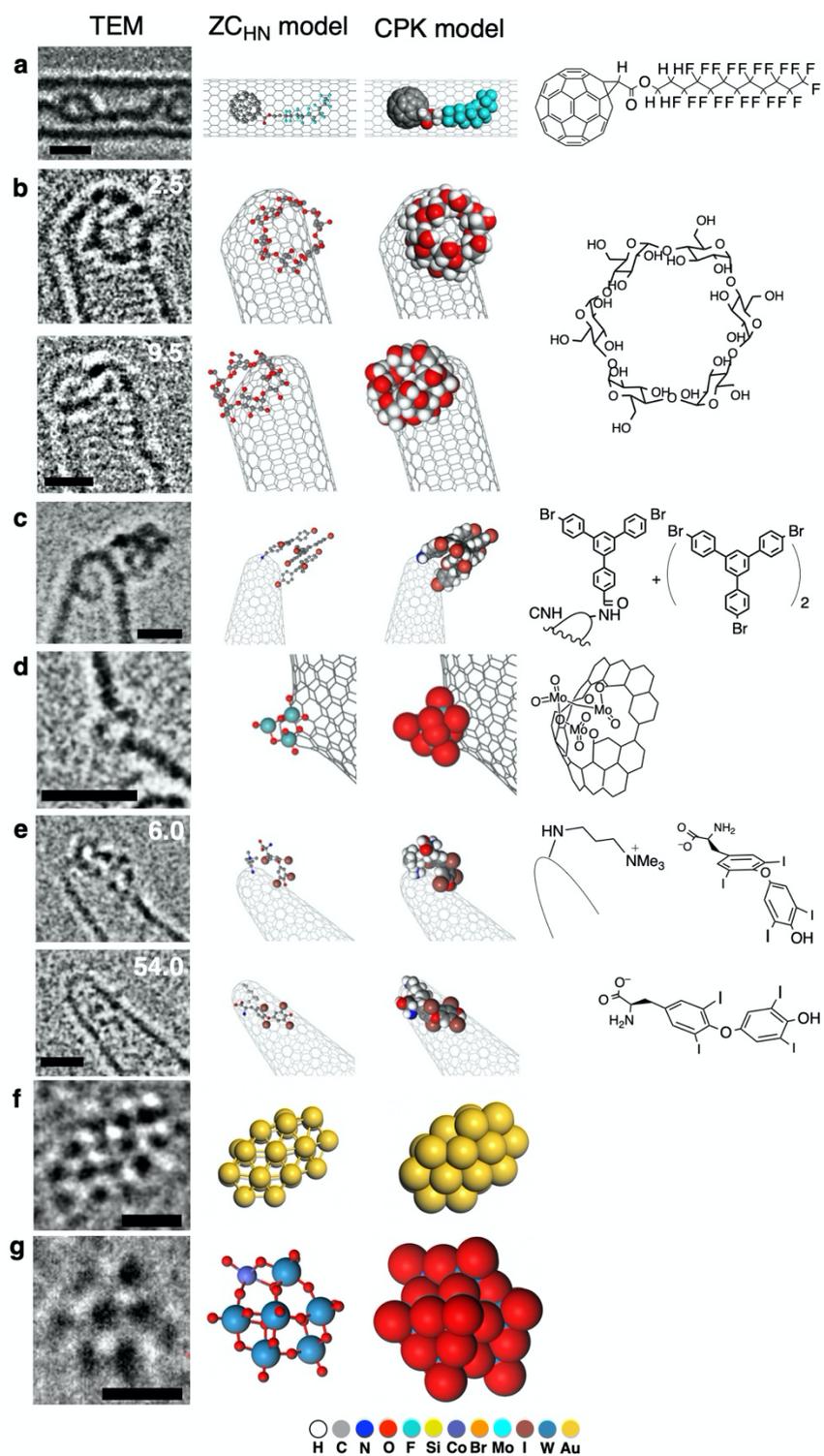

**Figure 5.** Experimental AR-TEM image and ZC$_{HN}$ model. (a) Perfluoroalkyl [60]fullerene in a single-walled CNT. See Figure S4 for a comparison of the TEM image and the ZC$_{HN}$ model image. Copyright 2014 American Chemical Society. (b)

Time evolution of TEM image of α-CD sliding on the CNH. Copyright 2021 American Chemical Society. (c) A termolecular van der Waals cluster on the CNH. Copyright 2012 Nature Publishing Group. (d) A trimolybdenum oxo cluster formed at an oxidized defect of the CNH. Copyright 2021 Chemical Society of Japan. (e) Time evolution of the TEM image of a thyroxine molecule changing its conformation and orientation on a CNH bearing a trimethylammonium ion linker. (f) A $Au_{24}$ cluster on an amorphous carbon film. Copyright 2021 American Chemical Society. (g) A $[CoSiW_{11}O_{39}]^{6-}$ cluster on graphene oxide. The silicon atom is hidden behind a tungsten atom. Copyright 2016 American Chemical Society. Scale bar: 1 nm for a-e and 0.5 nm for f and g.

The $ZC_{LN}$ parameter is useful for ultrathin crystalline samples as illustrated in Figure 1e for UiO-66,[15] which consists of a stable network made of $[Zr_6O_4(OH)_4]$ octahedron clusters and 1,4-benzene dicarboxylic acid (BDC) ligands. The TEM image in Figure 1e obtained by careful adjustment of parameters including contrast transfer function beautifully revealed the network structure, highlighted by the image of benzene rings in the BDC linker seen as benzene rings superimposed along the e-beam axis. Figure 1g shows the $ZC_{LN}$ model based on X-ray crystallographic data,[43] and we see the benzene ring balanced in size with the Zr atoms (sky blue). Figure 6a and 6b further illustrate application to a MOF, HKUST-1 made of dinuclear copper clusters connected by benzene-1,3,5-tricarboxylate linker molecules.[15] The $ZC_{LN}$ model made from a reported crystal structure[44] in Figure 6c highlights the Cu cluster (orange) seen clearly in the TEM image.

Figure 6d and inset show a TEM image of a crystal of MAPbBr$_3$ (MA = methylammonium ion) seen from the [100] axis.[15] Large and smaller dots corresponding to the atomic columns are arranged alternately in a grid pattern. By comparing the TEM image with the ZC$_{LN}$ model based on a reported X-ray crystal structure (Figure 6e),[45] we can assign the large dots to atomic columns consisting of alternating lead ($Z = 82$) and bromine ($Z = 35$), and the smaller dots to columns consisting of only bromine atoms. The lighter MA molecules are seen off the center of the void as smaller dots (indicated by arrows, Figure 6d inset), which are reproduced well in the ZC$_{LN}$ model (Figure 6e). In Figure 6f, we show a simulation without shot noise, where the lead/bromine and the bromine columns appear similar in size, which illustrates the importance of the inclusion of shot noise in TEM simulations.

Figure 6g shows a TEM image from the [100] axis of a NaCl nanocrystal in CNT.[6] The alternating overlap of sodium and chloride ions in the depth direction results in an array of uniformly sized dots, reflecting the contrast of the chlorine atoms of $Z = 17$ rather than the sodium atoms ($Z = 11$) (Figure 6h). This is also reproduced in the ZC$_{LN}$ model shown in the upper right of Figure 6g and 6i, where the size of the chlorine, but not the sodium, is reflected as the size of the atomic column. The atomic size is greatly underestimated when we use ZC$_{HN}$ radii (data not shown). We did not develop radius parameters for thick crystals, where interference among neighboring columns of atoms overwhelms the effects of interference caused by individual atoms.

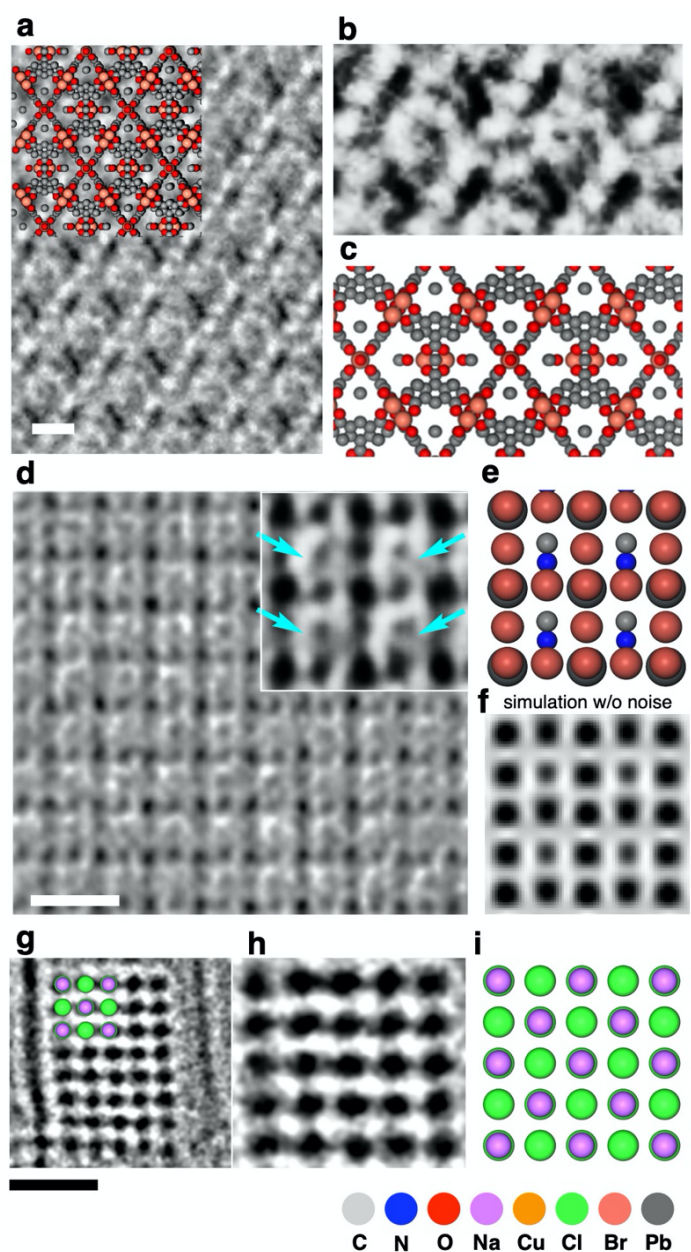

**Figure 6.** Experimental TEM image and ZC$_{LN}$ model for thin crystals of organic–inorganic hybrid materials and an ionic salt. (a) TEM image of HKUST-1. Copyright 2018 The American Association for the Advancement of Science. (b) A magnified view of a. (c) ZC$_{LN}$ model corresponding to b, which is also shown in the upper left corner in a. (d) TEM image of MAPbBr$_3$ and a magnified view (inset). Arrows indicate MA ions. Copyright 2018 The American Association for the Advancement of Science. (e) ZC$_{LN}$

model corresponding to d. (f) TEM simulation corresponding to e without shot noise. (g) TEM image of a nanocrystal of NaCl in CNT. Copyright 2021 American Chemical Society. (h) A magnified view of g. (i) $ZC_{LN}$ model corresponding to h, which is also shown in the upper left of g.

**Conclusion**

In summary, here we proposed ZC atomic radii for molecular modeling of AR-TEM images. Experience has taught us that heavier elements appear as both larger and more dense spots than lighter elements; though, in theory, their sizes should be essentially the same. We have shown that it is the shot noise that causes the Z-dependency of the image sizes of various elements, and parameterized the ZC atomic radii by balancing the atomic number and the effects of the shot noise inherent to digital imaging. We proposed two parameter sets, $ZC_{HN}$ and $ZC_{LN}$, for analysis of high-noise and low-noise AR-TEM images, and found that the former also works well for images where the substrate provides "noise" that interferes with the signal of the specimen being studied. We expect that the method will be further modified to be better as more cases are examined. We expect that the ZC model will accelerate research on visual molecular science as an increasing number of chemists gain access to AR-TEM instruments.

**Methods**

**TEM image simulation of a single atom.** TEM image simulation was carried out using elbis software by parallelized computation using a graphics processing unit.[28] Image simulations are generated in a 16-bit gray scale from an object exit wave function calculated with the multislice formalism (slice thickness: 0.010 nm). The atomic potential was calculated using inverse Fourier transformation from Weickenmeier–Kohl form factors.[46] The details of the calculation method are described in detail in previous papers.[28,35] Computational three-dimensional atomic models of single atoms of all elements were constructed in BIOVIA Materials Studio software. Atoms are located in the same plane (e.g., the same *z* value) in three-dimensional space with a separation of 20 Å, where a contrast from a single atom does not interfere with that of neighboring atoms. The parameters for simulation were set to be acceleration voltage = 80 kV, spherical aberration constant $C_s$ = –3 μm, pixel size = 0.01054 nm/pixel, defocus value = –10 nm (underfocus), which is a typical set of SMART-EM conditions at $2 \times 10^6$ magnification. TEM simulation images under a finite ED were generated based on the cumulative distribution of probability density[47] using an algorithm implemented in the software.

**ZC$_{HN}$ and ZC$_{LN}$ atomic radii.** The simulated TEM image of a single atom was processed using ImageJ 1.53c software to obtain the intensity profile. The intensity profile of the simulated atomic image was cut at two noise levels determined for the HN and LN conditions, and the length of the bottom of a triangle is defined as the ZC$_{HN}$ and ZC$_{LN}$ image sizes in Figure 3e. The HN and LN noise levels were defined as one standard deviation of the image intensity generated under the ED of $5.0 \times 10^5$ e$^-$ nm$^{-2}$ s$^{-1}$ for HN and $4.5 \times 10^6$ e$^-$ nm$^{-2}$ s$^{-1}$ for LN. We used these image sizes as the ZC$_{HN}$ and ZC$_{LN}$ atomic radii in the ZC molecular model.

**Acknowledgments**

We thank Profs. Kunio Takayanagi, Kaoru Yamanouchi, Andrei N. Khlobystov, Yu Han, and Dr. Fumio Hosokawa for discussion, Dr. Satori Kowashi for provision of experimental data, and Mr. Shingo Hasegawa for supporting TEM simulation. This research is supported by the Japan Society for the Promotion of Science (JSPS) KAKENHI (JP19H05459, JP20K15123, and JP21H01758) and the Japan Science and Technology Agency (CREST JPMJCR20B2). A part of this work was conducted in the Research Hub for Advanced Nano Characterization, The University of Tokyo and "Nanotechnology Platform" (project No. 12024046), both sponsored by MEXT. K.K. thanks JSPS for a predoctoral fellowship. J.X. and K.K. thank MEXT (ALPS program).

**Author Contributions** E.N. and K.H. supervised the project. J.X., K.T. and K.K carried out the TEM simulations and analyzed the data with T.N., K.H. and E.N. K.H. and E.N. cowrote the paper. All authors discussed the results and commented on the manuscript.

**Competing interests:** Authors declare no competing interests.

**Supplementary Information**

Supplementary Figures S1–S4.

**References**

[1] Brickmann, J.; Exner, T. E.; Keil, M.; Marhöfer, R. J. Molecular Graphics - Trends and Perspectives. *J. Mol. Model,* **2000**, *6*, 328–340.

[2] Ospadov, E.; Tao, J.; Staroverov, V. N.; Perdew, J. P. Visualizing Atomic Sizes and Molecular Shapes with the Classical Turning Surface of the Kohn–Sham Potential. *Proc. Natl. Acad. Sci.* **2018**, *115*, E11578–E11585.

[3] Corey, R. B.; Pauling, L. Molecular Models of Amino Acids, Peptides, and Proteins. *Rev. Sci. Instrum.* **1953**, *24*, 621–627.

[4] Shannon, R. D. Revised Effective Ionic Radii and Systematic Studies of Interatomic Distances in Halides and Chalcogenides. *Acta Crystallogr. A.* **1976**, *32*, 751–767.

[5] Okada, S.; Kowashi, S.; Schweighauser, L.; Yamanouchi, K.; Harano, K.; Nakamura, E. Direct Microscopic Analysis of Individual $C_{60}$ Dimerization Events: Kinetics and Mechanisms. *J. Am. Chem. Soc.* **2017**, *139*, 18281–18287.

[6] Nakamuro, T.; Sakakibara, M.; Nada, H.; Harano, K.; Nakamura, E. Capturing the Moment of Emergence of Crystal Nucleus from Disorder. *J. Am. Chem. Soc.* **2021**, *143*, 1763–1767.

[7] Hanayama, H.; Yamada, J.; Tomotsuka, I.; Harano, K.; Nakamura, E. Rim Binding of Cyclodextrins in Size-Sensitive Guest Recognition. *J. Am. Chem. Soc.* **2021**, *143*, 5786–5792.

[8] Harano, K. Self-Assembly Mechanism in Nucleation Processes of Molecular Crystalline Materials. *Bull. Chem. Soc. Jpn.* **2020**, *94*, 463–472.

[9] Koshino, M.; Tanaka, T.; Solin, N.; Suenaga, K.; Isobe, H.; Nakamura, E. Imaging of Single Organic Molecules in Motion. *Science* **2007**, *316*, 853.

[10] Nakamura, E. Atomic-Resolution Transmission Electron Microscopic Movies for Study of Organic Molecules, Assemblies, and Reactions: The First 10 Years of Development. *Acc. Chem. Res.* **2017**, *50*, 1281–1292.

[11] Skowron, S. T.; Chamberlain, T. W.; Biskupek, J.; Kaiser, U.; Besley, E.; Khlobystov, A. N. Chemical Reactions of Molecules Promoted and Simultaneously Imaged by the Electron Beam in Transmission Electron Microscopy. *Acc. Chem. Res.* **2017**, *50*, 1797–1807.

[12] Xing, J.; Schweighauser, L.; Okada, S.; Harano, K.; Nakamura, E. Atomistic Structures and Dynamics of Prenucleation Clusters in MOF-2 and MOF-5 Syntheses. *Nat. Commun.* **2019**, *10*, 3608.

[13] Kirkland, E. J. Atomic Size. In *Advanced Computing in Electron Microscopy* 2nd ed; Springer, 2010; pp 84–85.

[14] Zhu, Y.; Ciston, J.; Zheng, B.; Miao, X.; Czarnik, C.; Pan, Y.; Sougrat, R.; Lai, Z.; Hsiung, C.-E.; Yao, K.; Pinnau, I.; Pan, M.; Han, Y. Unravelling Surface and Interfacial Structures of a Metal–Organic Framework by Transmission Electron Microscopy. *Nat. Mater.* **2017**, *16*, 532–536.

[15] Zhang, D.; Zhu, Y.; Liu, L.; Ying, X.; Hsiung, C.-E.; Sougrat, R.; Li, K.; Han, Y. Atomic-Resolution Transmission Electron Microscopy of Electron Beam-Sensitive Crystalline Materials. *Science* **2018**, *359*, 675–679.

[16] Williams, D. B.; Carter, C. B. *Transmission Electron Microscopy: A Textbook for Materials Science*; 2nd Ed. ed.; Springer US: New York, 2009.

[17] Yamanouchi, K. Scattering Electrons. In *Quantum Mechanics of Molecular Structures*; Springer, 2013; pp 197–258.

[18] Shimizu, T.; Lungerich, D.; Stuckner, J.; Murayama, M.; Harano, K.; Nakamura, E. Real-Time Video Imaging of Mechanical Motions of a Single Molecular Shuttle with Sub-Millisecond Sub-Angstrom Precision. *Bull. Chem. Soc. Jpn.* **2020**, *93*, 1079–1085.

[19] Gross, L.; Mohn, F.; Moll, N.; Liljeroth, P.; Meyer, G. The Chemical Structure of a Molecule Resolved by Atomic Force Microscopy. *Science* **2009**, *325*, 1110–1114.

[20] Talirz, L.; Ruffieux, P.; Fasel, R. On-Surface Synthesis of Atomically Precise Graphene Nanoribbons. *Adv. Mater.* **2016**, *28*, 6222–6231.

[21] Lungerich, D.; Hoelzel, H.; Harano, K.; Jux, N.; Amsharov, K. Y.; Nakamura, E. A Singular Molecule-to-Molecule Transformation on Video: The Bottom-Up Synthesis of Fullerene $C_{60}$ from Truxene Derivative $C_{60}H_{30}$. *ACS Nano* **2021**, https://doi.org/10.1021/acsnano.1c02222.

[22] Markevich, A.; Kurasch, S.; Lehtinen, O.; Reimer, O.; Feng, X.; Müllen, K.; Turchanin, A.; Khlobystov, A. N.; Kaiser, U.; Besley, E. Electron Beam Controlled Covalent Attachment of Small Organic Molecules to Graphene. *Nanoscale* **2016**, *8*, 2711–2719.

[23] Nakamura, E. Movies of Molecular Motions and Reactions: The Single-Molecule, Real-Time Transmission Electron Microscope Imaging Technique, *Angew. Chem. Int. Ed*. **2013**, *52*, 236–252.

[24] Egerton, R. F. Measurement of Inelastic/Elastic Scattering Ratio for Fast Electrons and Its Use in the Study of Radiation Damage. *Phys. Status. Solidi.* **1976**, *37*, 663–668.

[25] Kirkland, E. J. Single Atom Images. In *Advanced Computing in Electron Microscopy* 2nd ed; Springer, 2010; pp 93–95.

[26] Koizumi, H.; Oshima, Y.; Kondo, Y.; Takayanagi, K. Quantitative High-Resolution Microscopy on a Suspended Chain of Gold Atoms. *Ultramicroscopy* **2001**, *88*, 17–24.

[27] Rullgård, H.; Öfverstedt, L. -G.; Masich, S.; Daneholt, B.; Öktem, O. Simulation of Transmission Electron Microscope Images of Biological Specimens. *J. Microscopy* **2011**, *243*, 234–256.

[28] Hosokawa, F.; Shinkawa, T.; Arai, Y.; Sannomiya, T. Benchmark Test of Accelerated Multi-Slice Simulation by GPGPU. *Ultramicroscopy* **2015**, *158*, 56–64.

[29] Cao, K.; Biskupek, J.; Stoppiello, C. T.; McSweeney, R. L.; Chamberlain, T. W.; Liu, Z.; Suenaga, K.; Skowron, S. T.; Besley, E.; Khlobystov, A. N.; Kaiser, U. Atomic Mechanism of Metal Crystal Nucleus Formation in a Single-Walled Carbon Nanotube. *Nat. Chem.* **2020**, *341*, 855.

[30] Koh, A. L.; Wang, S.; Ataca, C.; Grossman, J. C.; Sinclair, R.; Warner, J. H. Torsional Deformations in Subnanometer MoS Interconnecting Wires. *Nano Lett.* **2016**, *16*, 1210–1217.

[31] Chuvilin, A.; Khlobystov, A. N.; Obergfell, D.; Haluska, M.; Yang, S.; Roth, S.; Kaiser, U. Observations of Chemical Reactions at the Atomic Scale: Dynamics of Metal-Mediated Fullerene Coalescence and Nanotube Rupture. *Angew. Chem. Int. Ed.* **2010**, *49*, 193–196.

[32] Yuk, J. M.; Park, J.; Ercius, P.; Kim, K.; Hellebusch, D. J.; Crommie, M. F.; Lee, J. Y.; Zettl, A.; Alivisatos, A. P. High-Resolution EM of Colloidal Nanocrystal Growth Using Graphene Liquid Cells. *Science* **2012**, *336*, 61–64.

[33] Meyer, J. C.; Eder, F.; Kurasch, S.; Skakalova, V.; Kotakoski, J.; Park, H. J.; Roth, S.; Chuvilin, A.; Eyhusen, S.; Benner, G.; Krasheninnikov, A. V.; Kaiser, U. Accurate Measurement of Electron Beam Induced Displacement Cross Sections for Single-Layer Graphene. *Phys. Rev. Lett.* **2012**, *108*, 196102.

[34] Torabi, K.; Sayad, S.; Balke, S. T. Adaptive Image Thresholding for Real-time Particle Monitoring. *Int. J. Imag. Syst. Tech.* **2006**, *16*, 9–14.

[35] Gamm, B.; Blank, H.; Popescu, R.; Schneider, R.; Beyer, A.; Gölzhäuser, A.; Gerthsen, D. Quantitative High-Resolution Transmission Electron Microscopy of Single Atoms. *Microsc. Microanal.* **2012**, *18*, 212–217.

[36] Jia, C. L.; Mi, S. B.; Barthel, J.; Wang, D. W.; Dunin-Borkowski, R. E.; Urban, K. W.; Thust, A. Determination of the 3D Shape of a Nanoscale Crystal with Atomic Resolution from a Single Image. *Nat. Mater.* **2014**, *13*, 1044–1049.

[37] Nakamura, E.; Koshino, M.; Tanaka, T.; Niimi, Y.; Harano, K.; Nakamura, Y.; Isobe, H. Imaging of Conformational Changes of Biotinylated Triamide Molecules Covalently Bonded to a Carbon Nanotube Surface. *J. Am. Chem. Soc.* **2008**, *130*, 7808–7809.

[38] Harano, K.; Takenaga, S.; Okada, S.; Niimi, Y.; Yoshikai, N.; Isobe, H.; Suenaga, K.; Kataura, H.; Koshino, M.; Nakamura, E. Conformational Analysis of Single Perfluoroalkyl Chains by Single-Molecule Real-Time Transmission Electron Microscopic Imaging. *J. Am. Chem. Soc.* **2014**, *136*, 466–473.

[39] Harano, K.; Homma, T.; Niimi, Y.; Koshino, M.; Suenaga, K.; Leibler, L.; Nakamura, E. Heterogeneous Nucleation of Organic Crystals Mediated by Single-Molecule Templates, *Nat. Mater.* **2012**, *11*, 877–881.

[40] Kratish, Y.; Nakamuro, T.; Liu, Y.; Li, J.; Tomotsuka, I.; Harano, K.; Nakamura, E.; Marks, T. J. Synthesis and Characterization of a Well-Defined Carbon Nanohorn-Supported Molybdenum Dioxo Catalyst by SMART-EM Imaging. Surface Structure at the Atomic Level. *Bull. Chem. Soc. Jpn.* **2020**, *94*, 427–432.

[41] Hasegawa, S.; Takano, S.; Harano, K.; Tsukuda, T. New Magic $Au_{24}$ Cluster Stabilized by PVP: Selective Formation, Atomic Structure, and Oxidation Catalysis. *JACS Au* **2021**, *1*, 660–668.

[42] Barnard, A. W.; Zhang, M.; Wiederhecker, G. S.; Lipson, M.; McEuen, P. L. Real-Time Vibrations of a Carbon Nanotube. *Nature* **2019**, *566*, 89–93.

[43] Øien, S.; Wragg, D.; Reinsch, H.; Svelle, S.; Bordiga, S.; Lamberti, C.; Lillerud, K. P. Detailed Structure Analysis of Atomic Positions and Defects in Zirconium Metal–Organic Frameworks. *Cryst. Growth Des.* **2014**, *14*, 5370–5372.

[44] Yakovenko, A. A.; Reibenspies, J. H.; Bhuvanesh, N.; Zhou, H.-C. Generation and Applications of Structure Envelopes for Porous Metal–Organic Frameworks. *J. Appl. Crystallogr.* **2013**, *46*, 346–353.

[45] Wang, K.-H.; Li, L.-C.; Shellaiah, M.; Sun, K. W. Structural and Photophysical Properties of Methylammonium Lead Tribromide ($MAPbBr_3$) Single Crystals. *Sci. Rep.* **2017**, *7*, 13643.

[46] Weickenmeier, A.; Kohl, H. Computation of Absorptive Form Factors for High-Energy Electron Diffraction. *Acta Crystallogr. A* **1991**, *47*, 590–597.

[47] Gilbert, N.; Pollak, H. O. Amplitude Distribution of Shot Noise, *Bell Syst. Tech. J.* **1960**, *39*, 333–350.